\documentclass[aps,pre,twocolumn,superscriptaddress,amsmath,amssymb]{revtex4-1}
\usepackage{graphicx}
\usepackage{color}
\usepackage[EULERGREEK]{sansmath}

\def\longrightharpoonup{\relbar\joinrel\rightharpoonup}
\def\longleftharpoondown{\leftharpoondown\joinrel\relbar}

\def\longrightleftharpoons{
  \mathop{
    \vcenter{
      \hbox{
        \ooalign{
          \raise1pt\hbox{$\longrightharpoonup\joinrel$}\crcr
          \lower1pt\hbox{$\longleftharpoondown\joinrel$}
        }
      }
    }
  }
}

\newcommand{\rates}[2]{\displaystyle
  \mathrel{\longrightleftharpoons^{#1\mathstrut}_{#2}}}

\newcommand{\torate}[1]{\overset{#1}{\to}}

\newcommand*{\unit}[1]{\ensuremath{\mathrm{\,#1}}}

\newcommand{\beq}{\begin{equation}}
\newcommand{\beqs}{\begin{subequations}}
\newcommand{\beqnn}{\begin{equation*}}
\newcommand{\beqn}{\begin{eqnarray}}
\newcommand{\beqnnn}{\begin{eqnarray*}}
\newcommand{\bml}{\begin{multline}}
\newcommand{\eeq}{\end{equation}}
\newcommand{\eeqs}{\end{subequations}}
\newcommand{\eeqnn}{\end{equation*}}
\newcommand{\eeqn}{\end{eqnarray}}
\newcommand{\eeqnnn}{\end{eqnarray*}}
\newcommand{\eml}{\end{multline}}

\newcommand{\slabel}[1]{\label{sec:#1}}

\newcommand{\fref}[1]{Fig.~\ref{fig:#1}}

\newcommand{\flabel}[1]{\label{fig:#1}}
\newcommand{\eref}[1]{Eq.~\ref{eqn:#1}}
\newcommand{\erefs}[1]{Eqs.~\ref{eqn:#1}}
\newcommand{\elabel}[1]{\label{eqn:#1}}

\newcommand{\molsf}[1]{{\rm #1}}
\newcommand{\avg}[1]{\langle{#1}\rangle}
\newcommand{\avgin}[1]{\langle{#1}\rangle}

\newcommand{\br}[1]{\left ( #1\right )}
\newcommand{\brin}[1]{\left ( #1\right )}

\newcommand{\sqbr}[1]{\left [ #1\right ]}
\newcommand{\acco}[1]{\left \lbrace #1\right \rbrace}

\newcommand{\diff}[2]{\frac{d #1}{d #2}}

\newlength{\colwidth}
\newcommand{\capt}[2]{
\caption{\flabel{#1}
\begin{sansmath}\textsf{#2}\end{sansmath}}
}

\newlength{\figwidth}
\newcommand{\letter}[1] {
	\begin{minipage}[t]{0.01\figwidth}
		\vspace{0pt}
		\textsf{\small{#1}}
	\end{minipage}
}

\newcommand{\figwithletter}[3] {
\raggedright
	\letter{#1}
	\begin{minipage}[t]{#2\figwidth}
		\vspace{0pt}
		\includegraphics[]{#3}
	\end{minipage}
}

\begin{document}
\title{Multiplexing oscillatory biochemical signals}

\author{Wiet de Ronde}\affiliation{FOM Institute AMOLF, Science Park 104, 1098 XG Amsterdam, The
Netherlands}
\author{Pieter Rein ten Wolde}\affiliation{FOM Institute AMOLF, Science Park 104, 1098 XG Amsterdam,
The Netherlands}

\begin{abstract}
  In recent years it is increasingly being recognized that biochemical
  signals are not necessarily constant in time and that the temporal
  dynamics of a signal can be the information carrier. Moreover, it is
  now well established that components are often shared between
  signaling pathways. Here we show by mathematical modeling that
  living cells can multiplex a constant and an oscillatory signal:
  they can transmit these two signals through the same signaling
  pathway simultaneously, and yet respond to them specifically and
  reliably. We find that information transmission is reduced not only
  by noise arising from the intrinsic stochasticity of biochemical
  reactions, but also by crosstalk between the different
  channels. Yet, under biologically relevant conditions more than 2
  bits of information can be transmitted per channel, even when the
  two signals are transmitted simultaneously. These observations
  suggest that oscillatory signals are ideal for multiplexing signals.\\
\end{abstract}

\maketitle

\section{Introduction}
Cells live in a highly dynamic environment, which means that they
continually have to respond to a large number of different signals.
One possible strategy for signal transmission would be to use
distinct signal transduction pathways for the transmission of the respective
signals. However,  it
is now clear that components are often shared between different
pathways. Prominent examples are the Mitogen-Activated Protein Kinase
(MAPK) signaling pathways in yeast, which share multiple components
\cite{Schwartz2004,Rensing2009a}. In fact, cells can even transmit
different signals through one and the same pathway, and yet respond
specifically and reliably to each of them. Arguably the best-known
example is the rat PC-12 system, in which the epidermal growth factor
(EGF) and neuronal growth factor (NGF) stimuli are transmitted through
the same MAPK pathway, yet give rise to different cell fates, respectively
differentiation and proliferation \cite{Marshall1995,
  Sasagawa2005}. Another example is the p53 system, in which the
signals representing double-stranded and single-stranded breaks in the
DNA are transmitted via the same pathway \cite{Batchelor2011}.  These observations suggest
that cells can {\em multiplex} biochemical signals \cite{DeRonde2011},
{\em i.e.} transmit multiple signals through one and the same
signaling pathway, just as many telephone calls can be transmitted
simultaneously via a shared medium, a copper wire or the ether.

One of the key challenges in transmitting multiple signals via pathways that
share components is to avoid unwanted crosstalk between the
different signals. In recent years, several mechanisms for generating signaling
specificity have been proposed. One strategy is spatial insulation, in
which the shared components are recruited into distinct macromolecular
complexes on scaffold proteins \cite{Schwartz2004, Patterson2010}. This
mechanism effectively creates independent communication channels, one for each
signal to be transmitted. Another mechanism is kinetic insulation, in which the
common pathway is used at different times, and a temporal separation between the
respective signals is thus established \cite{Behar2007}.  Another
solution is cross-pathway inhibition, in which one signal dominates
the response \cite{Bardwell2007, McClean2007, Hu2009,Thalhauser2009,Hu2011}.
In the latter two schemes, kinetic insulation and cross-pathway inhibition, the
signals are effectively transmitted via one signaling pathway, though in these
schemes multiple messages cannot be transmitted simultaneously.

We have recently demonstrated that cells can truly multiplex signals:
they can transmit at least two signals simultaneously through a common
pathway, and yet respond specifically and reliably to each of them
\cite{DeRonde2011}. In the multiplexing scheme that we proposed, the
input signals are encoded in the concentration levels of the signaling
proteins. The underlying principle is, however, much more generic,
since essentially any coding scheme can be used to multiplex
signals. This observation is important, because it is becoming
increasingly clear that cells employ a wide range of coding strategies
for transducing signals. One is to encode the signals in the duration
of the signal.  This is the scheme used by the NGF-EGF system: while
EGF stimulation yields a transient response of ERK, NGF leads to a
sustained response of ERK \cite{Sasagawa2005,Marshall1995}. Another
strategy is to encode the message in the frequency or amplitude of
oscillatory signals. Indeed, a large number of systems have now been
identified that employ oscillatory signals to transduce
information. Arguably the best-known example is calcium oscillations
\cite{Berridge2000a}, but other examples are the p53
\cite{Batchelor2011}, NFAT \cite{Hoffmann2002, Nelson2004a}, nuclear
ERK oscillations \cite{Shankaran2009} and NF-$\kappa$B system
\cite{Hoffmann2002,Ashall2009b,Sung2009,Cheong2011}. In fact, cells use
oscillatory signals not only to transmit information intracellularly,
but also from one cell to the next---insulin \cite{Wu2011} and
Gonadotropin Release Hormone (GnRH) \cite{Li1989} are prominent examples
of extracellular signals that oscillate in time. More examples of
systems that encode stimuli in the temporal dynamics of the signal are
provided in the recent review article by Behar and Hoffmann
\cite{Behar2010}.

In this manuscript we demonstrate that oscillatory signals can be used
to multiplex biochemical signals. We present a multiplexing system,
which is based on well-known network motifs, such as the
Goldbeter-Koshland push-pull network \cite{Goldbeter1981a} and the incoherent feedforward motif \cite{Alon:2007tz}.  For the constant signal the
information is encoded in the concentration, while for the oscillatory
signal the message is encoded in the amplitude or the frequency of the
oscillations. These input signals are then multiplexed in the dynamics
of a common signaling pathway, which are finally decoded by downstream
networks. 

Our results highlight that information transmission is a
mapping problem. For optimal information transmission, each input
signal needs to be mapped onto a unique output signal, allowing the
cell to infer from the output what the input was. It is now well
established that noise, arising from the inherent stochasticity of
biochemical reactions, can reduce information transmission
\cite{Detwiler2000,Ziv2007, Tkacik2008, Walczak2009,
  Mehta2009,Tostevin2009, DeRonde2011, Cheong2011}, because a given output signal
may correspond to different input signals. Additionally, here we show that
crosstalk between the two different signals can also compromise
information transmission: a given state of a given input signal can
map onto different states of its corresponding output signal, because
the input-output mapping for that channel depends on the state of the
signal that is transmitted through the other channel.  This crosstalk presents a
fundamental bound on the amount of information that can be
transmitted, because it limits information transmission even in the
deterministic, mean-field limit. We also show, however, that under
biologically relevant conditions more than $2$ bits of
information can be transmitted per channel. We end by comparing our results with
observations on experimental systems, in which oscillatory and
constant signals are transmitted through a common pathway.

\section{Results}
\subsection{The model}
\fref{Fig1} shows a cartoon of the setup. We consider two
input species $\molsf{S}_1$ and $\molsf{S}_2$, with two corresponding
output species, $\molsf{X}_1$ and $\molsf{X}_2$, respectively.  The
concentration $S_1\br{t}$ of input $\molsf{S}_1$
oscillates in time, while the concentration of $\molsf{S}_2$ is
constant in time. An input signal can represent different messages;
that is, the input can be in different states.
 For $\molsf{S}_1$ the different states
could either be encoded in the amplitude or in the period of
the oscillations. Here, unless stated otherwise, we will focus on the
former and comment on the latter in the Discussion. 
The different states of $\molsf{S}_2$
correspond to different copy-number or, since we are working at
constant volume, concentration levels $S_2$. The
signals $S_1$ and $S_2$ drive
oscillations in the concentration $V(t)$ of an intermediate component
$\molsf{V}$, with a mean that is determined by
$S_2$ and an amplitude that is determined by $S_1$ (see {\em SI}). The
states of $\molsf{S}_1$ are thus encoded in the amplitude of $V(t)$ while the
states of $\molsf{S}_2$ are encoded in the mean level
of $V$. The output $X_2$ reads out the mean of $V(t)$ and hence the
state of its input $S_2$ by simply time-integrating the oscillations of
$V(t)$. The output $X_1$ reads out the amplitude of the oscillations
in $V(t)$ and hence the state of $S_1$ via an adaptive
network, indicated by the dashed circle. We now describe the coding and
decoding steps in more detail.

\subsubsection{Encoding}
In the encoding step of the motif, the two signals $S_1$, $S_2$ are combined
into the shared pathway. The signals are modeled as a 
sinusoidal function
\begin{align}
\elabel{sig}
S\br{t}=\mu\br{1+A\sin\br{2\pi \frac{t}{T}}}.
\end{align}
$\mu$ is the signal mean, $A$ is the signal amplitude  and $T$ is the period of the signal oscillation. We
assume that the signals are deterministic and discuss the effects of noise
later. As discussed above, $S_1$ is an
oscillatory signal, with kinetic parameters $A_1, T_1$ and constant $\mu_1$.
$S_2$ is constant, $A_2=0$, and the concentration level $\mu_2$ carries the
information (i.e. sets the state) in the signal. In recent years it has been shown
that biochemical systems can tune separately the amplitude and frequency of an
oscillatory signal \cite{Tsai2008a, Stricker2008}.

The simplest shared pathway is a single component,$\molsf{V}$,
which could be a receptor on the cell or nuclear membrane, but could
also be an intracellular enzyme or a gene regulatory protein. We
imagine that each signal
is a kinase for 
$\molsf{V}$, which can switch between an active (e.g. phosphorylated)
state ($\molsf{V}^{\molsf{P}}$) and an
inactive (e.g. unphosphorylated) state, such that
\begin{align}
\elabel{VPfull}
\diff{V^P}{t}=\frac{k_V\sqbr{\sum_i
S_i\br{t}}\br{V_T-V^P}}{K_V+\br{V_T-V^P}}-m_VE_T\frac{V^P}{M_V+V^P},
\end{align}
where we sum over the two signals $S_1\br{t}$ and $S_2\br{t}=\mu_2$. The dephosphorylation
is mediated by a phosphatase, that has a constant copy number $E_T$. In
\eref{VPfull} we assume Michaelis-Menten dynamics for $\molsf{V}$ (see
{\em SI} for more details).

The Michaelis-Menten kinetics of the activation of $\molsf{V}$ could
distort the oscillatory signal of $S_1$. They could reduce the
amplitude of the oscillations or transform the sinusoidal signal into a
 signal that effectively switches between two values. Such transformations potentially hamper
information transmission. We therefore require that the component $V$
accurately tracks the dynamics of the input signals. 
It is well known that a linear transfer function between $S$ and $V$ does not
lead to a deformation of the dynamic behavior, but only to a rescaling of the
absolute levels (see {\em SI}). A linear transfer function can be
realized if the kinase acts in the saturated regime, while the
phosphatase is not saturated ($K_V\ll \br{V_T-V^P\br{t}}, M_V\gg
V^P\br{t}$), leading to
\begin{align}
\elabel{VPlin}
\diff{V^P}{t}=k_V\br{\sum_i S_i\br{t}}-m'_VV^P.
\end{align}
with $m'_V=m_VE_T/M_V$.

\begin{figure*}[t]
	\begin{center}
	\begin{tabular}{c}
		\figwithletter{}{0.6}{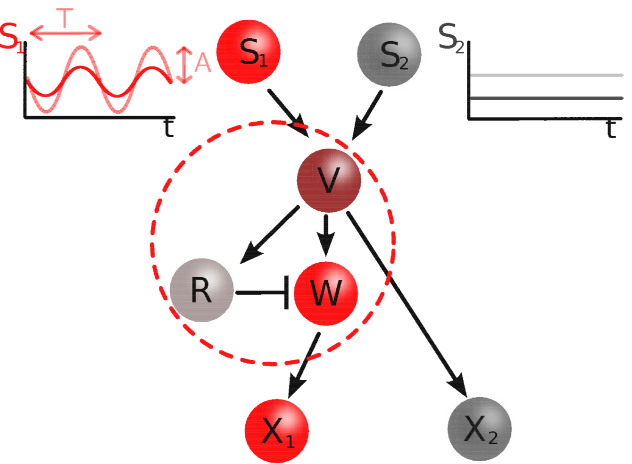}
	\end{tabular}
	\end{center}
	\capt{Fig1}{Schematic drawing of the multiplexing system. Two signals are
multiplexed. Signal $S_1$ oscillates in time while signal $S_2$ is constant. The
message of $S_1$ could either be encoded in the amplitude or the frequency of
the oscillations, but in this manuscript we focus on the former. The message of
$S_2$ is encoded in the concentration. The output or response of $S_1$ is
$X_1$ while the response of $S_2$ is $X_2$. Encircled is the
adaptive motif, used to readout the amplitude of the oscillations of $S_1$.
}
\end{figure*}

\subsubsection{Decoding $V^{P}$ to $X_1, X_2$}
The second part of the multiplexer is the decoding of the information
in $V^{\molsf{P}}$ into a functional output (see \fref{Fig1}).  The signals that are
encoded in $V^{\molsf{P}}$ have to be decoded into two output
signals, the responses $X_1$ and $X_2$. We imagine that the cell should be able
to infer from
an instantaneous measurement of the output response
the state of the corresponding input signal.
Therefore, we take the outputs of the multiplexing motif
to be the concentration levels $X_1$ and $X_2$ of the output species
$\molsf{X}_1$ and $\molsf{X}_2$, respectively. Here $X_1$ is
the response of ${S}_1$, while ${X}_2$ is the response of
${S}_2$.

The response ${X}_2$ should be sensitive to the concentration of
${S}_2$, but be blind to any characteristics of
${S}_1$. In our simple model there is only one time-dependent
signal, namely $S_1$; $S_2$ is constant in time.
Since $\molsf{V}^{\molsf{P}}$ has a linear transfer function of the
signals (\eref{VPlin}), the average level of $V^{\molsf{P}}$,
$\avgin{V^P}$, is independent of both the amplitude $A_1$
and the period $T_1$ of the oscillations in $S_1$. $\avgin{V^P}$
does depend on the mean concentration level of the two signals, and since
$S_1$ has a constant mean, changes in $\avgin{V^P}$ reflect only a
change in the mean of $S_2$, $\mu_2$. As a result, a simple
linear time integration motif can be used as the final read out for
$S_2$. We therefore model $X_2$ as
\begin{align}
	\elabel{x2}
	\diff{X_2}{t}=k_{X_2}V^P\br{t}-m_{X_2} X_2\br{t}.
\end{align}
Since \eref{x2} is linear, $\avgin{X_2}$ is a function of
$\avgin{V^P}$ only. Moreover, if the response time of $\molsf{X}_2$,
$\tau_{X_2}\br{=m^{-1}_{X_2}}$, is much longer than the oscillation
period $T_1$ of $S_1$, the effect of the oscillations on the
instantaneous concentration $X_2$ is integrated out. This is important
to reduce the variability in $\avgin{X_2}$ due to dynamics in the
system \cite{Tostevin2012}.

\begin{figure*}[h]
\begin{center}
	\figwithletter{a)}{0.9}{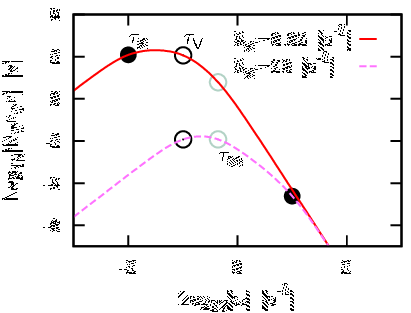}
	\figwithletter{b)}{0.9}{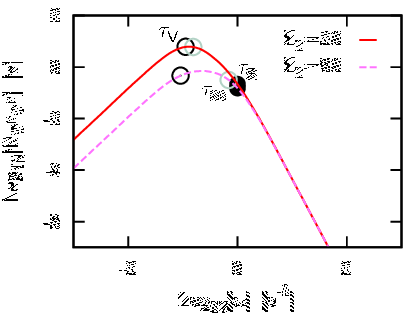}
\end{center}
\capt{Fig2}{The gain $g^2_{W^P}\br{\omega}$ for channel 1 for different
  parameter sets. The circles indicate the response times
  $\tau_i$. \textbf{a)} The effect of changing the timescale
  $k_R=\mu_R$. \textbf{b)} The effect of changing the signal $S_2$ in
  the other channel 2; it is seen that the gain $g^2_{W^P}\br{\omega}$
  of channel 1 depends on $S_2$, which may lead to
  crosstalk. Parameters: Panel \textbf{a)}: $\mu_2=500$,
  $k_R=m_R$. Panel \textbf{b)}: $\mu_2=200$ and $\mu_2=800$, $k_R=1,
  m_R=1$.  Panels \textbf{a,b)}: $\mu_1=0,
  k_V=0.1,K_V=10^{-4}V_T,m_VE_T=600,M_V=5V_T,V_T=1000$,$k_W=1,K_W=M_W=W_T/4,
  W_T=1000$, $m_W$ sets the timescale.
}
\end{figure*}

For $\molsf{X}_1$ a simple time-integration scheme does not work. The
information that has to be mapped onto the output concentration $X_1$
is the amplitude of $\molsf{S}_1$, which is propagated to $V^P$. The
output $X_1$ should therefore depend on the amplitude of the
oscillations of $V^P$, but not on its mean $\avgin{V^P}$, since the
mean represents the information in $\molsf{S}_2$. These requirements
mean that the frequency-dependent gain of the network from $V$ to
$X_1$ should have a band-pass structure. The frequency-dependent gain
shows how the amplification of the input signal depends on the
frequency of the signal \cite{DeRonde2010} (\fref{Fig2}). Due to the finite lifetime
of the molecules, the frequency-dependent gain of any biochemical
network inevitably reaches zero at high frequencies. Here we
require that at the other end of the frequency spectrum, in the
zero-frequency limit, the gain should also be small: Changes in the
constant level of $V_P$, which result from changes in $S_2$, should
not be amplified because $X_1$ should not respond to changes in
$S_2$. Indeed, only at intermediate frequencies should the gain be
large: Changes in the amplitude of the oscillations of $V_P$ at the
frequency of the input $S_1$ must be strongly amplified, because these
changes correspond to changes in $S_1$ to which $X_1$ must
respond. The network between $V$ and $X_1$ should thus have a
frequency-transmission band that matches the frequency of $S_1$.  The
output $X_1$ will then strongly respond to $S_1$ but not to $S_2$.

A common biochemical motif with a frequency band-pass filter is an
adaptive motif \cite{Tostevin2010a}.  An adaptive system does not
respond to very slowly varying signals, essentially because it then
already adapts to the changing signal before a response is
generated. Indeed, the key feature of an adaptive system is that the
steady-state output is independent of the magnitude of a constant
input, meaning that
\begin{align}
\avg{W}=f\br{\acco{\text{all parameters}}\notin \avg{V^P}}.
\end{align}
This appears to be precisely what is required, because it means
that when
$\molsf{S}_1$ is absent and {$S_2$} is changed, the output {$X_1$}
remains constant, as it should --- only {$X_2$} should change when
{$S_2$}, a signal constant in time, is changed. On the other hand,
while the steady-state output of an adaptive network is insensitive to
variations in constant inputs, it is sensitive to dynamical
inputs. This observation is well known; {it} is, e.g., the basis for
the chemotactic behavior of {\em E. coli}, where the system responds
to a change in the input concentration, and the strength of the
response depends on the magnitude of the change in input
concentration. This is another characteristic that {is required},
because it allows the magnitude of the response $X_1$ to depend on the
amplitude $A_1$ of the oscillations in $S_1$, thus enabling a mapping
from $A_1$ to $X_1$.  The important question that remains is
  whether the magnitude of the response solely depends on the change
  in the input concentration, which reflects $S_1$, or whether it also
  depends on the absolute value of the input concentration, which
  reflects $S_2$. In the following, we will show that it may depend on
  both, which would introduce crosstalk between the two signals.

Two common ways to construct an adaptive motif are known
\cite{Ma2009a}, the negative feedback motif and the incoherent
feed-forward motif. In this multiplexing system we use the latter. In the incoherent feed-forward
motif the input signal, in our case $V^P$, stimulates two downstream components
$\molsf{R,W}$, see \fref{Fig1}. One of the downstream components, $\molsf{R}$, is
also a signal for the other downstream component, $\molsf{W}$.
Importantly, the regulatory effect of the direct pathway
($V^P\to W$) is opposite to the effect of the indirect
pathway ($V^P\to R\to W$). As a result, if
$\molsf{V}^P$ activates $\molsf{W}$, this activation is counteracted by
the regulation of $\molsf{W}$ through $\molsf{R}$. We thus obtain
\begin{align}
\diff{R}{t}&=k_R V^P-m_R R, \elabel{eqr}\\
\diff{W}{t}&=k_{W} \frac{V^P\br{W_T-W^P}}{K_{W}+\br{W_T-W^P}}-m_{W} \frac{RW^P}{M_{W}+W^P}\elabel{eqy}.
\end{align}

This motif is adaptive, which can be shown by setting the
time-derivatives in \eref{eqr} and \eref{eqy} to zero and solving for the steady state
$\avgin{W^P}$. This yields
\begin{align}
\elabel{steady_state}
0&=\frac{\br{k_{W}\br{W_T-\avg{W^P}}}\br{m_R\br{\avg{W^P}+M_{W}}}}{\br{K_{W}+\br{W_T-\avg{W^P}}}\br{m_{W}k_R\avg{W^P}}}.
\end{align}
Although the full expression for $\avgin{W^P}$ is unwieldy to present,
\eref{steady_state} shows that it does not depend on the magnitude of a constant input $\avgin{V^P}$,
which means that the network is indeed adaptive.

For a correct separation of the signals, the response $W$ should be insensitive
to the average level of $V^P$, $\avgin{V^P}$, since $\avgin{V^P}$ carries
information {on} $\molsf{S}_2$, and not $\molsf{S}_1$. Indeed, a
dependence of $\molsf{W}$ on $\avgin{V^P}$ and hence on $\molsf{S}_2$
necessarily leads to unwanted crosstalk between the two information channels.
While the adaptive property of the network ensures that $W^{{P}}$ is
insensitive to the mean of ${V}^{{P}}$ for a constant input (see
\eref{steady_state}), this is not necessarily the case for a dynamic
$V^P\br{t}$. Since \eref{eqy} is non-linear, the response ${W}$ is dependent on
the precise functional form of ${V}^{{P}}$, and, more importantly, will depend on
$\avgin{V^P}$. Crosstalk
may thus arise, which will be studied in more detail below.

 The full expression of the frequency-dependent gain $g^2\br{\omega}$ is too
unwieldy to present here, but in simplified form we have
\begin{align}
\elabel{motif1_g2}
g^2_{W^P}\br{\omega}\propto \frac{\alpha
\omega^2}{\beta\br{\omega^2+\tau^{-2}_R}\br{\omega^2+\tau^{-2}_{W}}\br{
\omega^2+\tau^{-2}_V}},
\end{align}
where $\alpha$ and $\beta$ are proportionality constants and $\tau_i$ are the
response times of component $i$. For slowly varying signals
($\omega\to0$), the amplitude of the response is negligible due to the
$\omega^2$-term in the numerator of \eref{motif1_g2}, reflecting
the adaptive nature of the network. Second, for $\omega\ll
\min\sqbr{\tau^{-1}_V, \tau^{-1}_R, \tau^{-1}_{W}}$, the power scales with
$\omega^2$. For very large $\omega$ the power scales with $\omega^{-4}$. In the
intermediate regime for $\omega$, the scaling depends on the
precise response times. The response times are the diagonal Jacobian elements
for the linearized system  (\erefs{VPfull},\ref{eqn:eqr},\ref{eqn:eqy}),
\begin{align}
&\tau_{V}=\sqbr{\frac{m_VE_T
M_V}{\br{M_V+\avg{V^P}}^2}+\frac{k_VK_V\mu}{\br{V_T-\avg{V^P}+K_V}^2}}^{-1}
\elabel{tauv},\\
&\tau_{R}=m^{-1}_R\elabel{taur},\\
&\tau_{W}=\sqbr{\frac{m_{W}
M_{W}\avg{R}}{\br{M_{W}+\avg{W^P}}^2}+\frac{k_{W}K_{W}\avg{V^P}}{\br{W_T-\avg{
W^P}+K_{W}}^2}}^{-1}.\elabel{taux}
\end{align}
\eref{taur} gives the response time for a protein with a simple
birth-death reaction. The mathematical form of the response times,
$\tau_V$ and $\tau_W$, \eref{tauv} and \eref{taux}, resembles that of
a switching process with a forward and a backward step; their values
depend on the signal parameters. When the dynamics of $\molsf{V^P}$
operate in the linear regime (see \eref{VPlin}), $\tau_V$ simplifies
to $\tau_{V}\approx-\br{m_VE_T/\br{M_V}}^{-1}$, which is just the linear
decay rate of $\molsf{V}^{\molsf{P}}$.  Importantly, while $\avgin{W^P}$ is
independent of the state $\mu_2$ of $S_2$, the response time $\tau_W$
and hence the gain $g^2\br{\omega}$ 
do depend on $\avgin{V^P}$ and thereby on $S_2$. This means that the
response of $X_1$ to $S_1$ will depend on $S_2$, generating crosstalk.

The gain (\eref{motif1_g2}) is shown in \fref{Fig2}a,b for two
different parameter sets. The bandpass structure, with corresponding
resonance frequency (the peak in the gain) is observed. Further, with
circles, the response times $\tau_V$ (black open), $\tau_R$ (black
solid) and $\tau_W$ (gray open) are shown, which determine the
position of the peak in the gain; the peak occurs at a frequency in
between the two largest response times.  In \fref{Fig2}a we observe
the influence of increasing $k_R,m_R$, which are taken to be
equal. For very slow changes in ${R}$, corresponding to $k_R, m_R$
being {small}, the network has a very large gain. Increasing the
response time of ${R}$, decreases the amplitude at the resonance
frequency considerably. Faster tracking of ${V}^{{P}}$ by ${R}$ makes
the adaptation of the biochemical circuit very fast and as a result,
${W}^{{P}}$ does not respond at all to changes in ${V}^{{P}}$.  In
\fref{Fig2}b we observe the influence of changing the state $\mu_2$ of
$S_2$. The gain decreases for larger $\mu_2$, and the response time
$\tau_W$ increases. This may lead to crosstalk, since the mapping of
$A_1$ to $X_1$ will now depend on $S_2$.

Finally, we look at the last step in the motif, the conversion of the dynamic
response of the adaptive motif $W$ into ${X}_1$. The instantaneous concentration
$X_1$ should inform the system upon the {state} of input
{$\molsf{S}_1$}. Simple time-integration of ${W}$, similar to the
response ${X}_2$ (\eref{x2}), is not sufficient. While time-integration by
itself is important to average over multiple oscillation cycles, it is not
sufficient because time-integration with a linear-transfer function does not
lead to a change in the response when the amplitude of the input is varied,
assuming that the oscillations are symmetric.  Indeed to respond to
different amplitudes, a non-linear transfer function is required
\begin{align}
\elabel{eqx1}
\diff{X_1}{t}=k_{X_1}\frac{W^n}{W^n+K_{X_1}^n}-m_{X_1}X_1.
\end{align}
These Hill-type non-linear transfer functions are very common in biological
systems, for example in gene regulation by transcription factors, or protein
activation by multiple enzymes.

\subsection{Multiplexing}
\label{sec:ch8:results}
Now that we have specified the model with its components, we
characterize its multiplexing capacity, using the formalism of
information theory \cite{Shannon1948}. We
define two measures: 1) $I_1\br{X_1;A_1}$, the mutual information
between the concentration $X_1$ and the amplitude $A_1$ of signal
$\molsf{S}_1$, and 2) $I_2\br{X_2;\mu_2}$, the mutual information between
the concentration $X_2$ and the concentration level $\mu_2$ of
$\molsf{S}_2$.  The information capacity of the system is then defined
by the total information $I_{T}=I_1\br{X_1;A_1}+I_2\br{X_2;\mu_2}$
that is transmitted through the system. The mutual information in bits
shows how many different input states can be transmitted with
100\% fidelity. It does not necessarily reflect whether all
input signals are transmitted reliably. For example, increasing the number
of input states $N_A$ can increase the mutual information
$I_1\br{A_1;X_1}$ \cite{Shannon1948}, yet a specific output
concentration $X_1$ could be less informative about a specific input
amplitude $A_1$. To quantify the fidelity of signal transmission, we
 normalize the mutual information by the information
entropy $H(A_1)$ and $H(\mu_2)$ of the respective inputs. We therefore define the
relative mutual information
\begin{align}
\elabel{I_R}
I_R\br{\br{A_1;X_1},\br{\mu_2;X_2}}&=\frac{I_1\br{X_1;A_1}}{H\br{A_1}}+\frac{
I_2\br{X_2;\mu_2}}{H\br{\mu_2}}\\
&=\frac{I_1\br{X_1;A_1}}{\log_2\sqbr{N_A}}+\frac{I_2\br{X_2;\mu_2}}{\log_2\sqbr{
N_\mu}}.
\end{align}
Note that $I_R\brin{\brin{A_1;X_1},\brin{\mu_2;X_2}}$ has a maximum
value of $2$, meaning that each channel $i=1,2$ transmits all its
messages $S_i\to X_i$ with $100\%$ fidelity. 

The mutual information depends on the kinetic parameters of the
system, on the input distribution of the signal states, and on the
amount of noise in the system. In a previous study we have shown that
under biologically relevant conditions, a simple biochemical system
using only constant signals is capable of
simultaneously transmitting at least two bits of information
\cite{DeRonde2011}, meaning that at least two signals with two input
states can be transmitted with 100\% fidelity. Here we wondered
whether this information capacity can be increased. Therefore, we
study the system for increasing number of input states (increasing
$N_A$ for $S_1$ and $N_\mu$ for $S_2$), where we assume a uniform
distribution of the states for $\molsf{S}_1$, $A_1\in \sqbr{0:1}$, and
for $\molsf{S}_2$, $\mu_2 \in \sqbr{0:\mu_{\rm max}}$ (see
\eref{sig}). To obtain a lower bound on the information that can be
transmitted, we optimize the total mutual information
 over a subset of the kinetic parameters,
where we constrain
the kinetic rates between $10^{-3}< k_i <10^3$, the
dissociation constants between $1<K_i<7.5\times 10^4$, 
the maximum concentration level for $S_2$ $10<\mu_{\rm max}<1000$ and the oscillation period $10<T<10000$.
We set the response times of $\molsf{X}_1, \molsf{X}_2$ to be much
longer than the oscillation period, so that the variability in $V$ and
$W$ due to the oscillations in $S_1$ is time-integrated; specifically,
$m_{X_1}=m_{X_2}=\br{NT_p}^{-1}{\rm s}^{-1}$, such that the output
averages over $N=10$ oscillations with period $T_p$. The noise
strength is calculated using the linear-noise approximation \cite{VanKampen2007}
while
assuming that the input signals are constant, of magnitude
$\mu_1,\mu_2$. The effects of the non-linear and oscillatory nature of
the network on the noise strength are thus not
taken into account. However, we do not expect that these two effects
qualitatively change the observations discussed below. To compute the noise
strength, we assume that the maximum copy numbers of $X_1$ and
$X_2$ are $1000$.
The optimization is performed using an evolutionary algorithm (see {\em SI}).

Before we discuss the information transmission capacity of our system,
we first show typical results for the time traces and input-output
relations as obtained by the evolutionary algorithm. \fref{Fig3}a
shows that the oscillations in $V_P$ are amplified by the adaptive
network to yield large amplitude oscillations in $W^P$. In contrast,
$X_1$ and $X_2$ only exhibit very weak oscillations due to their long
lifetime. \fref{Fig3}b shows that when $A_1$ is increased while
$\mu_2$ is kept constant, the average of $V^P$, which is set by
$\mu_2$, is indeed constant. As a result, $\avg{X_2}$ is constant, as
it should be (because $\mu_2$ is constant). In contrast, $X_1$
increases with $A_1$. This is because the amplitude of the
oscillations in $W^P$ increases with $A_1$, which is picked up by the
non-linear transfer function from $W^P$ to $X_1$. In addition, $X_1$
increases because the mean of $W^P$ itself increases, due to the
non-linearity of the network; this further helps to
increase $X_1$ with $A_1$. \fref{Fig3}c shows that when $\mu_2$ is
increased while $A_1$ is kept constant, $\avg{V^P}$ and hence $X_2$---the response of $S_2$---increases. Importantly, while the mean of the
buffer node $R$ of the adaptive network increases with $\avg{V^P}$,
the mean of the output of this network, $W^P$, is almost
constant. Consequently, $X_1$ is nearly constant, as it should
because $X_1$ should reflect the value of $A_1$ which is kept
constant. These two panels thus show that this system can multiplex
two signals: it can transmit multiple states of two signals through
one and the same signaling pathway, and yet each output responds very
specifically to changes in its corresponding input. This is the
central result of our manuscript.  

Interestingly, \fref{Fig3}c shows a (very) weak dependence of $X_1$ on
$S_2=\mu_2$, which will introduce crosstalk in the system. It is important
to realize that this will
reduce information transmission, even in a deterministic noiseless
system. The mechanism by which crosstalk can reduce information transmission
is illustrated in \fref{Fig4}. Maximal information transmission
between $S_1$ and $X_1$ occurs if a given amplitude $A_1$ (independent of
$\mu_2$) uniquely
maps to a specific output $X_1$, and similarly for $S_2$ and $X_2$. In
a deterministic system, every combination of inputs $(S_1,S_2)$ maps
onto a unique combination of outputs $(X_1,X_2)$. We aim to multiplex
the signals such that $X_1$ should be a function of $S_1$ (i.e. $A_1$)
only, while $X_2$ should be a function of $S_2$ ($\mu_2$) only. That
is, the mapping from $S_i$ to $X_i$ should be independent of the state
of the other signal $S_{j\neq i}$. However, crosstalk causes the
mapping from $S_i$ to $X_i$ to depend on the state of the other
channel. This dependence reduces information transmission, because a
given concentration of $X_1$ can now correspond to multiple values of
$A_1$, as illustrated in \fref{Fig4}a. Crosstalk can thus reduce
information transmission even in a deterministic system without
biochemical noise.

\begin{figure*}[t]
\begin{center}
\begin{tabular}{ccc}
	\figwithletter{a)}{0.5}{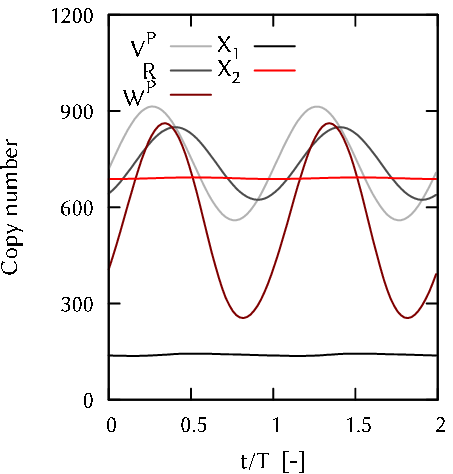}
&
	\figwithletter{b)}{0.5}{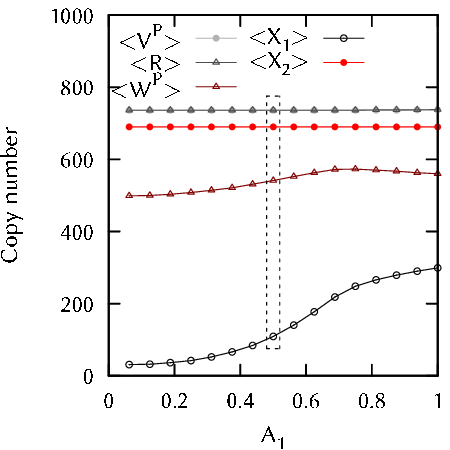}
&
	\figwithletter{c)}{0.5}{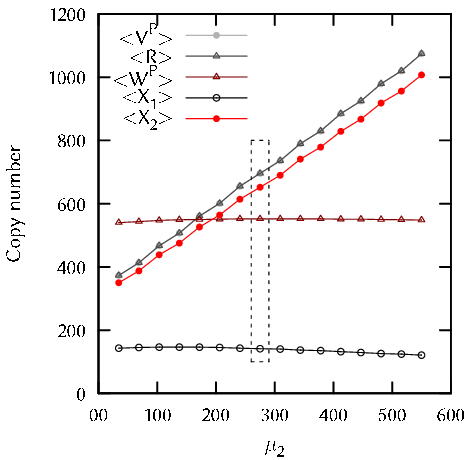}
\end{tabular}
\end{center}
\capt{Fig3}{Typical time traces and input-output curves as obtained by the
  evolutionary algorithm. Shown are results for a system with
  $N_A=N_{\mu}=16$.  \textbf{a)} Time traces of $V^P$, $R$, $W^P$,
  $X_1$ and $X_2$ for $A_1=0.5$ and $\mu_2=275$ \textbf {b)}
  $\avgin{V^P}$, $\avgin{R}$, $\avgin{W^P}$, $\avgin{X_1}$, and
  $\avgin{X_2}$ as a function of $A_1$, keeping $\mu_2=275$
  constant. \textbf{c)} $\avgin{V^P}$, $\avgin{R}$, $\avgin{W^P}$,
  $\avgin{X_1}$, and $\avgin{X_2}$ as a function of $\mu_2$, keeping
  $A_1=0.5$.  The figure shows that the system can multiplex: $X_1$ is
  sensitive to $S_1=A_1$ (panel b) but not $S_2=\mu_2$ (panel c),
  while $X_2$ is sensitive to $S_2=\mu_2$ (panel c) but not $S_1=A_1$ (panel
  b). The time traces in panel a correspond to the points in panels b
  and c that are indicated by the dashed lines. All panels correspond
  to the point in \fref{Fig5}c.
}
\end{figure*}

It is of interest to quantify the amount of information that can be
transmitted in the presence of crosstalk in a deterministic, noiseless
system. Via the procedure described in the {\em SI}, we compute the
maximal mutual information for the two channels, assuming that we have
a uniform distribution of input states for each channel, with $A_1\in
\sqbr{0:1}$ and
$\mu_2 \in \sqbr{0:\mu_{\rm max}}$. We find that for channel $2$,
the mutual information is given by the entropy of the input
distribution, which means that the number of signals that can be
transmitted with 100\% fidelity through that channel is just the
total number of input signals for that channel. This is because signal
transmission through channel $2$ is hardly affected by crosstalk from
the other channel. Below we will see and explain that this observation
also holds in the presence of biochemical noise. For signal
transmission through channel $1$, however, the situation is markedly
different. The maximum amount of information that can be transmitted
through that channel is limited to about 4 bits. This means that up to
$2^4$ signals can be transmitted with 100\% fidelity; in this regime,
the input signal $S_1$ can be uniquely
inferred from the output signal $X_1$. Increasing the number of input
signals beyond $2^4$, however, does
not increase the amount of information that is transmitted through
that channel; more signals will be transmitted, but, due to the
crosstalk from the other channel, each signal will
be transmitted less reliably (see \fref{Fig4} and {\em SI}).
 
We will now quantify how many messages can be transmitted reliably in
the presence of not only crosstalk, but also biochemical noise. 
The results of the optimization of the mutual
information using the evolutionary algorithm are shown in \fref{Fig5}. The left panel shows the relative mutual information
for channel $1$, the middle panel for channel $2$, and the right panel
shows the total relative mutual information (\eref{I_R}). Clearly, biochemical noise affects information
transmission through the two respective channels differently. 

Firstly, we see that  the fidelity of signal transmission through channel $2$ is
effectively independent of the number of states $N_A$ that are transmitted
through channel $1$, even in the presence of biochemical noise (\fref{Fig5}b).
This means that channel $2$ is essentially insensitive to crosstalk
from channel $1$. This is because $X_2$ time-integrates the
sinusoidal $V_P(t)$ via a linear transfer function---the output $X_2$ is thus
sensitive to the mean of $V_P$ (set by $S_2$), but not to the amplitude
of $V_P(t)$ (set by $S_1$). We also observe that even in the presence of noise, the relative
information stays close to 100\% when $N_{\mu}$ is below $3$
bits. Channel $2$ is thus fairly resilient to biochemical noise, which
can be understood by noting that a linear transfer function (from
$V_p$ to $X_2$) allows for an optimal separation of the $N_{\mu}$ input
states in phase space \cite{Tkacik2009,Tkacik2011,Tkacik2012}. 

The left channel, $\molsf{S}_1 \to \molsf{X}_1$, is more susceptible
to noise (\fref{Fig5}a) and to crosstalk from the other channel,
$\molsf{S}_2$. The susceptibility to noise can be seen for $N_\mu=1
{\rm bit}=2$ states: the relative information decreases as $N_A$
increases. This sensitivity to noise becomes more pronounced as
$N_\mu$ increases, an effect that is due to the crosstalk from the
other channel. A larger $N_\mu$ reduces the accessible phase space for
channel $1$---it reduces the volume of state space that allows for a
unique mapping from $S_1$ to $X_1$. As a result, a
small noise source is more likely to cause a reduction in
$I_R(S_1;X_1)$. How crosstalk and noise together reduce information
transmission is further elucidated in \fref{Fig4}c. Remarkably, even in
the presence of noise, maximal relative information is obtained for
$N_A=N_\mu=4 (=2{\rm bits})$ (\fref{Fig5}c), showing that $4$ input
states can be transmitted for each channel simultaneously without loss
of information.
\begin{figure*}[t]
\begin{center}
\begin{tabular}{ccc}
	\figwithletter{a)}{0.58}{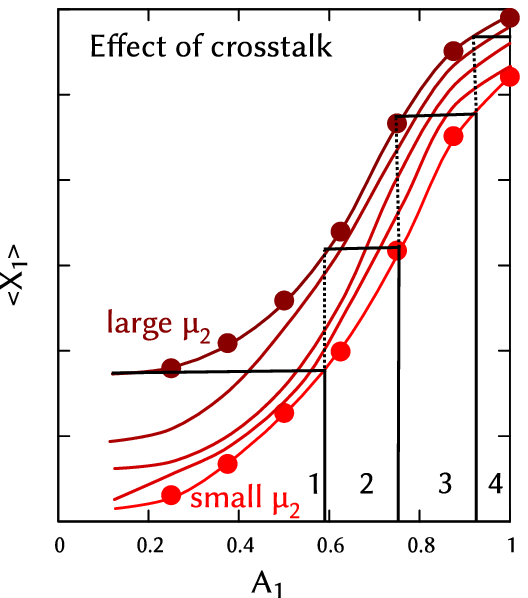}&
	\figwithletter{b)}{0.58}{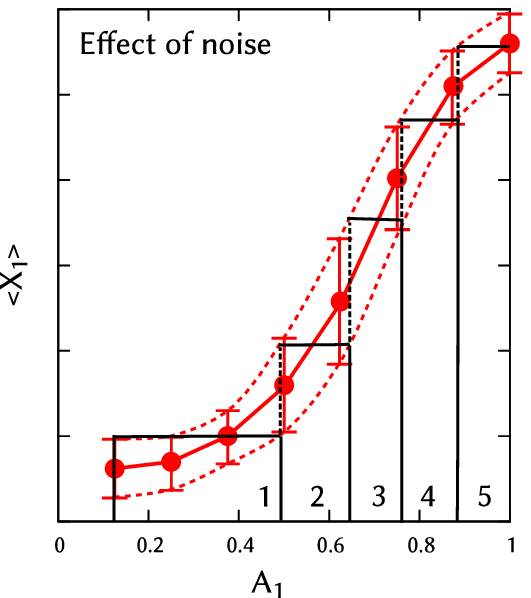}&
	\figwithletter{c)}{0.58}{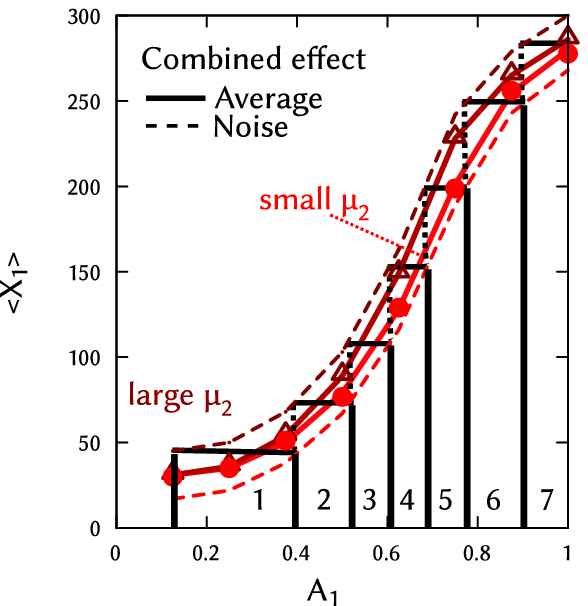}
\end{tabular}
\end{center}
\capt{Fig4}{The influence of noise and crosstalk on information
  transmission in pathway $S_1 \to X_1$.  \textbf{a)} Schematic: Crosstalk
  reduces the amount of information that can be transmitted. For every
  $A_1$ multiple values of $\avgin{X_1}$ are obtained, each
  corresponding to a different value of $\mu_2$. The dark red
  corresponds to the maximum value of $\avg{X_1}$ for each $A_1$,
  while the light red line denotes the minimum value. The black line
  in between the red lines visualizes the range for which a specific
  $\avgin{X_1}$ uniquely maps to a single input amplitude
  $A_1$. Crosstalk from the $\molsf{S}_2\to\molsf{X}_2$ channel thus
  limits the number of states, and hence the amount of information,
  that can be transmitted through channel $1$.  \textbf{b)} Schematic:
  Also noise
  reduces the number of input states that can be resolved.  Shown is
  the mean response curve $\avgin{X_1}(A_1)$ together with the noise
  in $X_1$.  Dotted lines give the minimum and maximum values of $X_1$
  for each amplitude. Since for each $A_1$ a larger range of $X_1$
  values is obtained, less states $A_1$ can be uniquely encoded in the
  phase space. This is reflected in the width of the boxes; indeed,
  here only $5$ input states can be transmitted with absolute
  reliability.  \textbf{c)} Combined effect of noise and crosstalk on
  information transmission for a system with $N_A=N_{\mu}=8$, as
  obtained from the evolutionary optimization algorithm; the results
  corresponds to the black dot in \fref{Fig5}c. Both the noise and the
  crosstalk reduce the number of possible input states that can be
  transmitted. Solid lines give the deterministic dose-response curve,
  while dashed lines correspond to a network with noise. Dark red
  lines indicate the maximum of $\avgin{X_1}$ for a specific $A_1$
  over the range of possible values of $\mu_2$, while red lines give
  the minimum value. Because for each $A_1$ a range of $\avgin{X_1}$
  values is obtained, the number of states $A_1$ that can be uniquely
  encoded in the phase space is limited. This is reflected in the
  increase in the width of the boxes; indeed, here only $7$ input
  states can be transmitted with absolute reliability.
}
\end{figure*}

\subsection{Experimental observations}
Here we connect our work to two biological systems. The first system is the
p53 DNA damage response system. The p53 protein is a cellular signal for
DNA-damage. Different forms of DNA damage exist and they lead to different
temporal profiles of the p53 concentration. Double-stranded breaks cause
oscillations in the p53 concentration, while single stranded damage leads to a
sustained p53 response \cite{Fei2003, Sengupta2005,Batchelor2011}. Compared to
our simple multiplexing motif, the encoding scheme in this system is more
involved. In our system two external signals activate the shared component
$\molsf{V}$. In the p53 system, p53 itself is $\molsf{V}$, but
interestingly, negative (indirect) autoregulation of p53 is required to obtain
sustained oscillations. 

Although the encoding structure is different, the main result is that the
system is able to encode two different signals into different
temporal profiles simultaneously; depending on the type of damage either a
constant and/or an oscillatory profile of p53 is present. These
two signals could therefore be transmitted simultaneously due to their
difference in the temporal profiles. 
For the p53 system the input signals are binary, e.g. either there is DNA
damage or not, although some experiments suggest that the amount
of damage also could be transmitted \cite{Lahav2004}. The maximum information
that can be transmitted following our simplified model is much
larger than that required for two binary signals. A mathematical model, based
upon experimental observations, shows that the encoding step
creates a temporal profile for p53 that could be decoded by our suggested
decoding module (not shown).

Another system of interest is the MAPK (or RAF-MEK-ERK) signaling
cascade. The final output of this cascade is the protein ERK, which
shuttles between the cytoplasm and the nucleus. ERK is regulated by
many different incoming signals of which EGF, NGF and HRG are well
known \cite{Shaul2007}. The temporal profile of ERK depends on the
specific input that is present. NGF and HRG lead to a sustained ERK
level \cite{Sasagawa2005}, while EGF leads to a transient or even
oscillatory profile of the ERK level \cite{Kholodenko2000,
  Sasagawa2005, Shankaran2009}. In comparison with our model ERK would
be the shared component $\molsf{V}$. Experiments show that
oscillations in the ERK concentration can arise due to intrinsic
dynamics of the system \cite{Shankaran2009}. However, these
oscillations could be amplified by, or even arise because of,
oscillations in the signal EGF, especially since, to our knowledge, it
is unclear what the temporal behavior of EGF is under physiological
conditions. 

For both experimental systems, we have only described the encoding step. In
both cases, two signals are encoded in a shared component
$\molsf{V}$, where one signal leads to a constant response, while the other
signal creates oscillations. Both p53 and ERK are transcription
factors for many downstream genes \cite{VonKriegsheim2009, Wei2006a}. For the
decoding of the constant signal, only a simple birth-death process
driven by $\molsf{V}$ would be required. Many genes are regulated in
this way \cite{Alon:2007tz}.
The decoding of the oscillatory signal requires an adaptive motif.
Although adaptive motifs are common in biological processes \cite{Alon:2007tz}, it is unclear
whether downstream of either p53 or ERK an adaptive motif is present, which
would complete our suggested multiplexing motif. As such, our study
should be regarded as a proof-of-principle demonstration that
biochemical networks can multiplex oscillatory signals.

\begin{figure*}[!ht]
	\begin{center}
\begin{tabular}{c|c|cl}
	$I_R\br{A_1,X_1}$ & $I_R\br{\mu_2,X_2}$ &
$I_R\br{\br{A_1,X_1},\br{\mu_2,X_2}}$&\\
\hline
\figwithletter{a)}{0.5}
{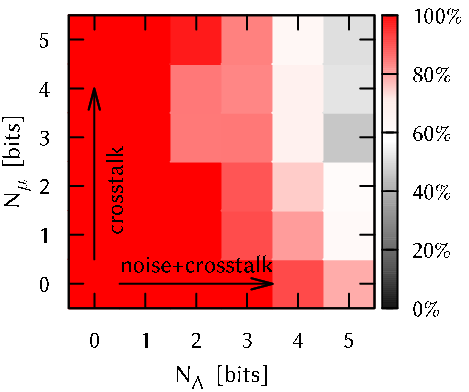}
&
\figwithletter{b)}{0.5}
{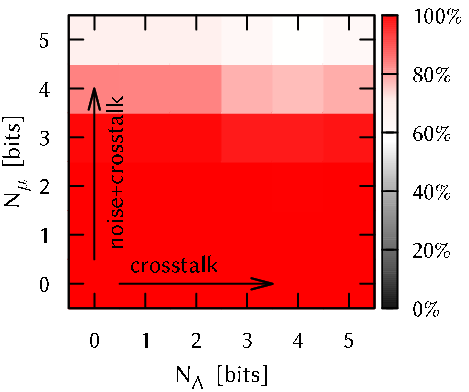}
&
\figwithletter{c)}{0.5}
{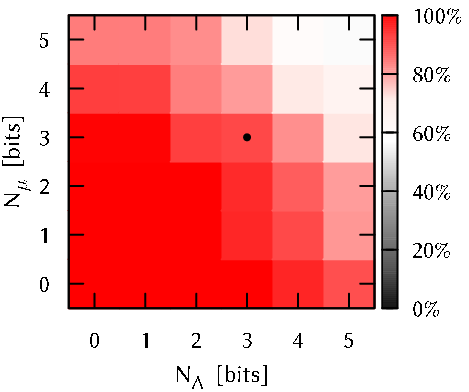}&
\end{tabular}
\end{center}
\capt{Fig5}{The transmitted relative information $I_R$ (\eref{I_R}) as
  function of the number of input states $N_{A}, N_{\mu}$, where $2$
  bits correspond to $2^2=4$ input states. Results are shown for a
  stochastic system with $X_T=1000$. In panels \textbf{a,b}
  100\% corresponds to $I_R=1$, while in \textbf{c} 100\%
  corresponds to $I_R=2$.   \textbf{a)} the relative mutual information
  $I_R\br{A_1,X_1}$ for the $\molsf{S}_1\to\molsf{X}_1$ channel; the total
mutual information is
  obtained by multiplying $I_R$ with $\log_2\brin{N_A}$, the
  horizontal axis.  Both the decrease in $I_R\brin{A_1,X_1}$ as a
  function of $N_A$ due to the presence of biochemical noise and the decrease in
$I_R\br{A_1,X_1}$ as a
  function of $N_\mu$ due to the presence of
  crosstalk is observed. \textbf{b)} the relative mutual
  information $I_R\br{\mu_2,X_2}$ for the $\molsf{S}_2\to\molsf{X}_2$
  channel. The total mutual
  information is obtained by multiplying $I_R$ with
  $\log_2\brin{N_\mu}$, the vertical axis. The effect of noise is relatively
  small and crosstalk from $S_1$ is hardly
  present.   \textbf{c)} the
  relative information of the total network
  $I_R\br{\br{A_1,X_1},\br{\mu_2,X_2}}=I_R\br{A_1,X_1}+I_R\br{\mu_2,X_2}$. The
  dot correponds to the timetraces in \fref{Fig3}.
 All results are obtained
  through numerical optimization (see {\em SI}).
}
\end{figure*}

\section{Discussion}
We have presented a scheme for multiplexing two biochemical
signals. The premise of the proposal is that the two signals have to
be transmitted, not integrated. Indeed, the central hypothesis is that
$X_1$ should only respond to $S_1$ and $X_2$ only to
$S_2$. Information transmission is then maximized when the crosstalk
between the different channels is minimized.

The model discussed here consists of elementary motifs, and  can
simultaneously transmit two signals reliably. One of these signals is
constant in time,
and its corresponding information is encoded in its concentration level, while
the other signal is dynamic, and its information is encoded in the dynamical
properties, but not in its average concentration level. The decoding of the
constant signals is performed by a time-integration motif, while the decoding
of the oscillatory signal requires a frequency-sensitive motif, for example an
adaptive motif.

The main problem in multiplexing biochemical signals is crosstalk
between the two signals. In this system the signals are encoded based
upon their dynamical profile---${S}_1$ is oscillatory and ${S}_2$ is
constant in time. The decoding module for the oscillatory signal, an
adaptive motif, is non-linear. Therefore, this motif is 
sensitive not only to the temporal properties like the amplitude, but also
to the mean or average of its input. This inevitably leads
to crosstalk between channel $1$ and channel $2$, reducing information
transmission.

Remarkably, the system is capable of transmitting over $3$ bits of
information through each channel with 100\% fidelity. In the presence
of noise the information transmission decreases, but even with
considerable noise levels in the biologically relevant regime, more
than $2$ bits of information can be transmitted through each channel
simultaneously; this information transmission capacity is comparable
to what has been measured recently in the context of NF-$\kappa$B signaling \cite{Cheong2011}. To transmit signals without errors it is preferable to
send most information using channel $2$ and a smaller number of states
through channel $1$. The reason for this is twofold. First, channel $2$ is
less noisy since the number of components is smaller; secondly,
channel $1$ is corrupted by crosstalk from channel $2$, leading to
overlaps in the state space of $X_1$ as a function of $A_1$ (see
\fref{Fig4}). Nonetheless, the two channels can reliably transmit 4
states in
the presence of noise. This is a considerable increase in
the information transmission compared to a system where both signals
are constant in time \cite{DeRonde2011}---this could transmit two
binary signals with absolute fidelity. This indicates that oscillatory
signals could significantly enhance the information transmission
capacity of biochemical systems. Importantly, while we have optimized
the parameters of our model system using an evolutionary algorithm, it
is conceivable that other architectures than those studied here allow
for larger information transmission. Indeed, the results presented
here provide a lower bound on information transmission.  

In this system we have assumed that the amplitude of the oscillatory
signal is the information carrier of that signal. The same analysis
could be performed for an oscillatory signal at constant amplitude but
with different frequencies. Qualitatively, the results will be
similar.  The dependence of the gain on the frequency 
means that the amplitude of the output varies with the frequency of
the input (see
\fref{Fig2}). The amplitude of the output thus characterizes the
signal frequency. However, an intrinsic redundancy is present in using
the frequency as the information carrier, which can be understood from
the symmetry of the gain (see \fref{Fig2}). The response of the system
is equal for frequencies that are positioned symmetrically with
respect to the resonance frequency. As a result, for any given output,
there are always two possible input frequencies, and without
additional information, the cell can not resolve which of the two
frequencies is present. Of course, one way to avoid this, would be to
use only a part of the gain, in which the gain increases monotonically
with frequency.

In this study we have assumed that the input signals are deterministic.
Results are obtained following deterministic simulations, where
noise is added following a solution of the linear-noise approximation assuming
non-oscillatory inputs. The effect of noise is a reduction of
the information transmission. However, the effect of noise can always be
counteracted by increasing the copy number. At the cost of producing
and maintaining more proteins, similar results can therefore be obtained
\cite{DeRonde2011}. The effect of oscillations on the
variability of the output is small since the response times of $\molsf{
X}_1$ and $\molsf{X}_2$ are much longer than the oscillation period. Slower
responding outputs would time-average the oscillation cycles even more, reducing
the variability in the response further.

Transmitting information via oscillatory signals has many advantages.
Oscillatory signals minimize the prolonged exposure to high levels of
the signal, which can be toxic for cells, as has been argued for
calcium oscillations \cite{Trump1995}. In systems with cooperativity
\cite{Gall2000b}, an oscillating signal effectively reduces the signal
threshold for response activation. Pulsed signals also provide a way
of controlling the relative expression of different genes
\cite{Cai2008}. Encoding of stimuli into oscillatory signals can
reduce the impact of noise in the input signal and during signal
propagation \cite{Rapp1981}. Frequency encoded signals can be decoded
more reliably than constant signals \cite{Tostevin2012}.  

Here we show that information can be encoded in the amplitude or
frequency of oscillatory signals, which are then decoded using a
non-linear integration motif.  We also discussed two biological
systems that may have implemented this multiplexing strategy. The idea
to use the temporal kinetics as the information carrier in a signal
has been studied in a slightly different context, where the dose
information is encoded in the duration of an intermediate component,
which in turn is time-integrated by a downstream component
\cite{Behar2008}.   Here, we show that encoding
signals into the temporal dynamics of a signaling pathway allows for
multiplexing, making it possible to simultaneously transmit multiple
input signals through a common network with high fidelity. It is
intruiging that systems with a bow-tie structure, such as calcium and
NF-$\kappa$B \cite{Cheong2011}, tend to transmit information via
oscillatory signals.

\section{Materials and Methods}
The model is based on mean-field chemical rate equations or the
linear-noise approximation \cite{Elf2003}. For details see the
Supporting Information.

\begin{acknowledgments}
  We thank Jos\'{e} Alvarado for a critical reading of the manuscript.
\end{acknowledgments}

\clearpage
\newpage

\begin{center}
{\bf \large Supplementary information: Multiplexing oscillatory biochemical
signals}\\[0.2cm]
Wiet de Ronde and Pieter Rein ten Wolde
\end{center}

\section{General definitions}
We use the following two general definitions for the mean and the maximum of a
specific component, if the period of the input signal is $T_p$
\begin{align}
\avg{Z}&=\frac{1}{T_p}\int_t^{t+T_p} Z\br{t'}dt',\\
Z_{\rm max}&=\sup_{T_p} Z\br{t}.
\end{align}

\subsection{Encoding}
\subsubsection{\slabel{MM_approx}MM-approximation}
In the derivation of Eq.~3 of the main text, we have assumed, as is
commonly done, Michaelis-Menten (MM) kinetics.  However, the
MM-approximation may not hold for a dynamical system
\cite{Segel1988a,RamiTzafriri2007a}, since the MM-approach is a coarse
graining of the full mass-action kinetics of the system.  Therefore we
numerically evaluate whether the MM-approximation is valid for our
network. The full set of reactions is (see Eq.~2 of the main text)
\begin{align}
\elabel{mass_act}
\molsf{S} + \molsf{V} &\rates{k_1}{k_{-1}} \molsf{SV}, \nonumber\\
\molsf{SV} &\torate{k_2} \molsf{S} + \molsf{V}^\molsf{P},\nonumber\\
\molsf{V}^\molsf{P} + \molsf{E} &\rates{m_1}{m_{-1}}
\molsf{V}^\molsf{P}\molsf{E},\nonumber\\
\molsf{V}^\molsf{P}\molsf{E} &\torate{m_2} \molsf{V}+\molsf{E}.
\end{align}
The quasi-steady-state approximation in MM-kinetics assumes that the
concentration of the complex of enzyme and substrate is in equilibrium and does
not change (or changes very slowly) with time, leading to
\begin{align}
\diff{SV}{t}=0,\,\diff{V^PE}{t}=0.
\end{align}
We compare the results for the full system (described by
\eref{mass_act}), with the Michaelis-Menten approximation in the
linear regime (Eq.~2 of the main text), in which the phosphorylation
reaction is linear in the signal concentration $S$ (\fref{si:Fig1}a).
Eq. 2 of the main text is obtained for
$K_V\br{=\br{k_{-1}+k_2}/k_{1}}\ll V_T-V^P$, such that the
phosphorylation reaction of $\molsf{V}$ to $\molsf{V}^\molsf{P}$
simplifies to $k_V\sum_i S_i\br{t}$ (Eq. 3 of the main text), which is indeed
linear in $S_{i}$ and zero-order in $V$.  If $V_T\gg S$, the
linearity condition will also be fulfilled for
all moments in time when the system is dynamical \cite{Segel1988a}.  In
this case all $\molsf{S}$ directly binds to $\molsf{V}$ and the
complex $\molsf{SV}$ is very stable.

This very stable complex $\molsf{SV}$, however, is likely to influence
the dynamics of the signal oscillations, as in the following two
examples: if the oscillations in the signal $\molsf{S}$ are driven due
to factors external of the cell (like hormone pulsing), or if the
oscillations depend on the (saturated) degradation of the signal
$\molsf{S}$ \cite{Mengel2010a} and this regulation does not act on the
signal in bound form. Due to the stability of the complex, in the
rising part of the oscillation $\molsf{S}$ directly forms the complex
$\molsf{SV}$, but, again because of the stability, during the falling
part of the oscillation $\molsf{SV}$ does not dissociate. Since the
complex is not regulated, the total signal level $S_T=S+SV$ increases,
since with every oscillation the concentration $\molsf{SV}$ increases,
until all $\molsf{V}$ is saturated. As a result, the oscillatory
dynamics of the signal is strongly influenced: the periodicity is
reduced and the mean level of signal present is increased
(\fref{si:Fig1}b).

The interaction between the signal $\molsf{S}$ and $\molsf{V}$ may corrupt the
information that is
encoded in the oscillations (\fref{si:Fig1}b), because
it influences the oscillation characteristics. To overcome this,
instead of assuming $S\ll V$, one could assume $V\ll S$. However, in
this regime it is unclear whether the dynamical behavior of the MM-approximation
accurately represents the dynamics of the full mass-action equations. Moreover,
if $S\gg V$, a small minimum concentration $S_{\rm min}$ directly saturates all
the $\molsf{V}$ molecules. As a result, the concentration of $\molsf{V}$ is
insensitive to any oscillation in the concentration $S$ when the minimum
concentration $S_{\rm min}$ is larger than the concentration $V$.

Therefore, an extended model is required, with two additional reactions
\eref{mass_act_2}. This extension is biologically inspired, since many
external signals are sensed by receptors $\molsf{Q}$, which in turn activate (or
phosphorylate) intracellular proteins. The crucial ingredient is that the
signal-bound receptor dissociates on a much faster timescale than the
oscillations. Due to this very fast receptor dissociation, the signal-bound
state is very small.

\begin{align}
\elabel{mass_act_2}
\molsf{S} + \molsf{Q} &\rates{q_1}{q_{-1}} \molsf{Q}^\molsf{A}, \nonumber\\
\molsf{Q}^\molsf{A} + V &\rates{k_1}{k_{-1}} \molsf{VQ}^\molsf{A}, \nonumber\\
\molsf{VQ}^\molsf{A} &\torate{k_2} \molsf{V}^\molsf{P} + \molsf{Q}^\molsf{A},
\nonumber\\
\molsf{V}^\molsf{P} + \molsf{E} &\rates{m_1}{m_{-1}}
\molsf{V}^\molsf{P}\molsf{E},\nonumber\\
\molsf{V}^\molsf{P}\molsf{E} &\torate{m_2} \molsf{V}+\molsf{E},
\end{align}

With this biological-inspired, small extension, the dose-response curve is
similar to the dose-response curve for the Michaelis-Menten approximation with
small $K_{V}$ (in the linear regime), also for oscillatory signals
(\fref{si:Fig1}c).

\begin{figure*}[!ht]
\begin{center}
		\figwithletter{a)}{0.6}{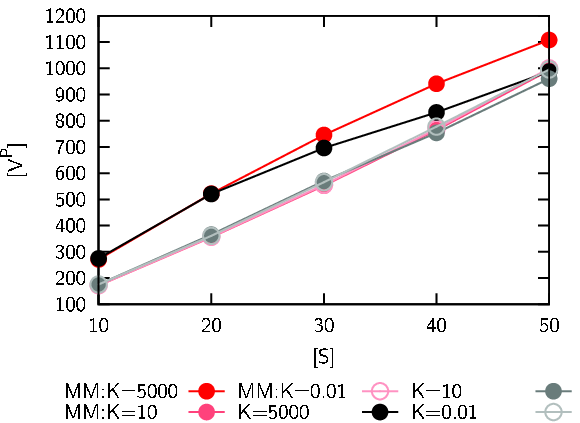}
		\figwithletter{b)}{0.6}{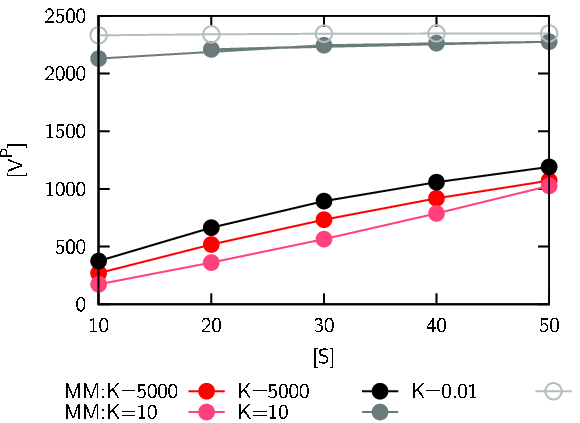}
		\figwithletter{c)}{0.6}{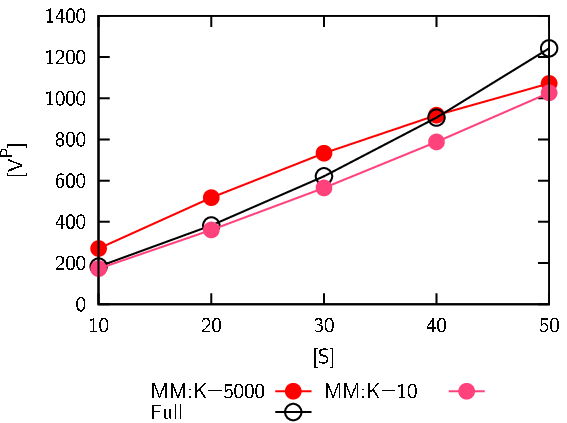}
\end{center}
\capt{si:Fig1}{Comparison of the results of the Michaelis-Menten
  approximation (red) with those of the full mass action equation
  (\eref{mass_act}, black) using Gillespie simulations.  \textbf{a)}
  Dose-response curve for {\em constant} signal $S$. The Michaelis
  Menten approximation describes the full system (\eref{mass_act})
  very well (only for $K=5000$ the curves do not precisely overlap).
  \textbf{b)} Dose-response for {\em sinusoidal} input $S$ with
  $T=100\unit{s}$. Due to the strong complex formation of $\molsf{VS}$
  all the signaling molecules $S$ bind $V$ until all $V$ is saturated
  and, as a result, the signal does not oscillate independently of the
  system. Accordingly, the effective concentration $S$ is much larger
  than the mean of the oscillations.  Parameters: $r_1\unit{s_{-1}},
  E_T=150,V_T=2500,M_V=5000$, and for $K_V=5000, K_V=10, K_V=0.01$
  respectively $\acco{k_{-1}, k_{2},m_{-1},m_{2}}=\acco{4997.5, 2.5,
    4999,1}\unit{s^{-1}},\acco{9, 1,4998,2}\unit{s^{-1}},\acco{0,
    1,4998, 2}\unit{s^{-1}}$, and $k_1\avg{S}=10,
  10,1000\unit{s^{-1}}$ respectively \textbf{c)} The dose-response
  curve of the extended model described by \eref{mass_act_2} compared
  to the Michaelis-Menten equations of Eq. 3 of the main text. It is
  seen that the MM model of Eq. 3 of the main text accurately
  describes the dynamics of the extended system of
  Eq. \eref{mass_act_2}.  Parameters: $q_1\avg{S}=2.5\unit{s^{-1}},
  q_{-1}=4000\unit{s^{-1}},k_1V_T=5000\unit{s^{-1}}, k_{-1}=4000,
  k_2=25\unit{s^{-1}}, m_1E_T=100\unit{s^{-1}}, ,
  m_{-1}=4998\unit{s^{-1}},m_2=1\unit{s^{-1}}, E_T=50, R_T=1000,
  V_T=5000$. }
\end{figure*}

\subsection{\slabel{lin_approx}Linear Approximation}
In this section we show that the MM-approximation in the linear regime
(see Eq. 3 of the main text) does not change the mean of
$\molsf{V}^\molsf{P}$, $\avgin{V^P}$, irrespective of the signal
characteristics. We compare analytical results with numerical
simulations and for completeness we compare this linear regime (I)
with two other regimes which we describe in more detail below:
zero-order dynamics for $\molsf{V}$ (II) and non-linear phosphorylation
(III).

In regime II there are many more $\molsf{V_T}$ molecules than kinase
($\molsf{S}_i$) and phosphatase ($\molsf{E_T}$) molecules. Therefore,
the kinase and phosphatase enzymes are {\em both} saturated
$\br{V_T\to\infty}$. Since saturation of both kinase and
phosphatase molecules implies that $M_V, K_V\ll V_T$,  the dynamics
of Eq. 2 of the main text can be simplified to
\begin{align}
\elabel{VP2sat}
\diff{V^P}{t}\approx k_V\br{\sum_i S_i\br{t}}-m_V.
\end{align}
This is the well-known regime of zero-order dynamics for $\molsf{V}$. In this
regime $\avgin{V^P}$ can be approximated by a binary function
$\avgin{V^P}=0$ or $\avgin{V^P}=V_T$, where the transition occurs at a
critical kinase concentration $\br{\sum_i S_i}_{\rm crit}$ (see \fref{si:Fig2},
open black symbols). $\molsf{V}$ thus acts as a switch. A
switch-like functional dependence of $\molsf{V}$ on $\molsf{S}$ does not lead to
perfect tracking of the signal $\molsf{S}$, and therefore not to accurate
propagation of the oscillations.

The third regime, regime III, is in a sense the opposite of the
previous. In this regime, $\molsf{V}$ is limiting (e.g. $M_V, K_V\gg
V_T$, see Eq. 2 of the main text). Eq. 2 of the main text then reduces
to
\begin{align}
\elabel{VP2lin}
\diff{V^P}{t}=k'_V\br{\sum_i S_i\br{t}}\br{V_T-V^P}-m'_VV^P,
\end{align}
where $k'_V=k_V/K_V$ and $m'_V=m_VE_T/M_V$. In this regime, the
phosphorylation and dephosphorylation reaction are first order in $V$
and $V^P$, respectively. A typical
dose-response curve is shown in \fref{si:Fig2} (closed gray symbols).

In regime I, corresponding to Eq. 3 of the main text,  the two preceding regimes
are combined  (\fref{si:Fig2}, closed red symbols). There is saturation of the
kinases in the production, but
saturation of $\molsf{V}^\molsf{P}$ in the dephosphorylation, leading to
\begin{align}
\elabel{SI:VPlin}
\diff{V^P}{t}=k_V\br{\sum_i S_i\br{t}}-m'_VV^P,
\end{align}
which is Eq. 3 of the main text.

In \fref{si:Fig2}a the dose-response curve for $\molsf{V}^\molsf{P}$
as function of $\molsf{S}$ is shown; the focus is on the mean
$\avgin{V^P}$ as a function of a constant signal $\molsf{S}$ with
increasing mean $\mu$.  Regime I of the main text (closed red symbols)
has an approximate linear relation between $S$ and
$\avgin{V^P}$. Regime II (open black symbols) shows
the switch-like response, while regime III (closed gray symbols) increases
hyperbolically to saturation.

A sinusoidal oscillation can only be propagated perfectly as a
sinusoidal signal if
the dose-response function is linear. For non-linear dose-response functions,
oscillations with small amplitude are propagated correctly, since for small
perturbations every function has linear characteristics. However, larger
amplitude oscillations are deformed by the non-linear transfer function. As a
result, the mean $\avgin{V^P}$ changes as a function of $A$ and/or $T$, the
oscillation parameters.
A stronger non-linear dose-response function decreases the amplitude-range of
oscillations that can be propagated without this deformation.
Figure~\ref{fig:si:Fig2}b shows $\avgin{V^P}$ for signals with different
properties $A$ and $T$, but constant signal mean $\mu$. If the transfer function
allows for perfect tracking of the input signal,  $\avgin{V^P}$ should
be constant because the mean $\mu$ is
constant. Indeed, in
both regime I and III $\avgin{V^P}$ does not depend on the oscillation
parameters $A, T$, while in regime II a strong dependence on these parameters
exist.

For completeness, in \fref{Fig3SI}a-c we show corresponding time traces. Again
the strong non-linear response for regime II (\fref{Fig3SI}b) is observed, while
regime I (\fref{Fig3SI}a) and III (\fref{Fig3SI}c) exhibit oscillations that
are
very similar to the sinusoidal oscillations of the input signal. Please also
note the
reduction in amplitude in regime III (\fref{Fig3SI}c), compared to
regime I of the main text (\fref{Fig3SI}a). This can be
explained
by the hyperbolic shape of the dose-response curve for regime III, which dampens
changes in $S$ (\fref{si:Fig2}a).

\begin{figure*}[!ht]
	\begin{center}
		\figwithletter{a)}{0.6}{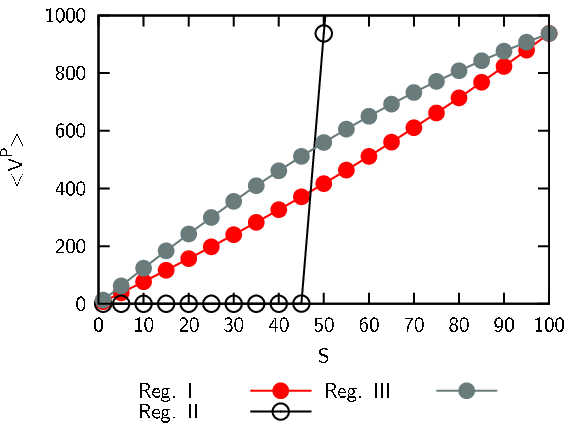}
		\figwithletter{b)}{0.6}{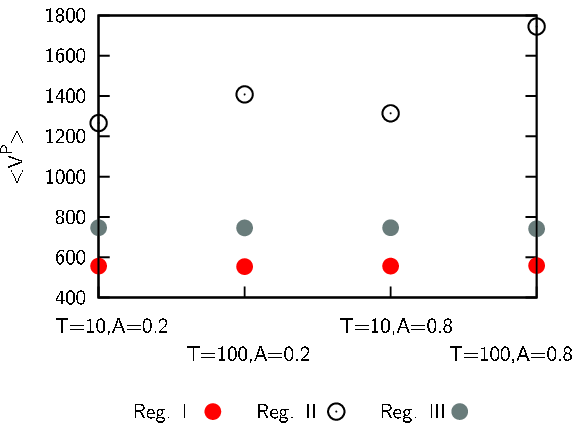}
	\end{center}
        \capt{si:Fig2}{ \textbf{a)} The dose-response curve is shown
          for the regime in which both the production and degradation
          are zero-order in $\molsf{V}^\molsf{P}$ (regime II,
          \eref{VP2sat}), in which both production and degradation are
          first order in $\molsf{V}^\molsf{P}$ (regime III,
          \eref{SI:VPlin}), and, the model of the main text, zero-order
          production but linear degradation of $\molsf{V}^\molsf{P}$
          (regime I,\eref{VP2lin}). The curve that is linear over the
          widest $S$-range is that for linear degradation of
          $\molsf{V}^\molsf{P}$, but zero-order production, studied in
          the main text (regime I). Parameters: Regime I:
          $m_VE_T=500$, $K_V/V_T=1/25000$, $M_V/V_T=2$;
          $k_V=1\unit{s^{-1}}$; Regime II: $m_VE_T=50$,
          $K_V/V_T=1/25000$, $M_V/V_T=1/25000$; Regime III:
          $m_VE_T=100$, $K_V/V_T=2$, $M_V/V_T=2$; \textbf{b)} The time
          average over a single oscillation period, $\avgin{V^P}_T$,
          is shown for four different simulations where the signal
          characteristics are as indicated and $\mu_{S}=50$.}
\end{figure*}

\begin{figure*}[!ht]
	\begin{center}
		\figwithletter{a)}{0.6}{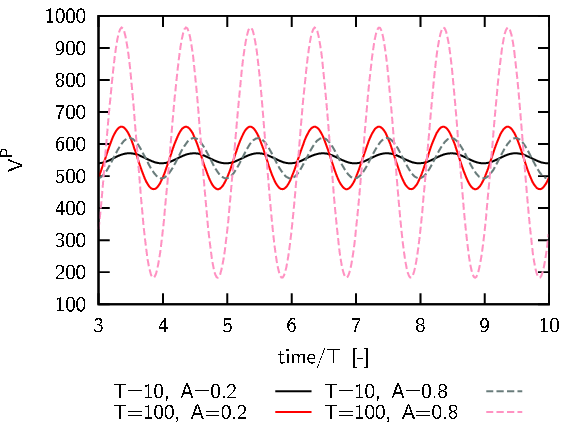}
		\figwithletter{b)}{0.6}{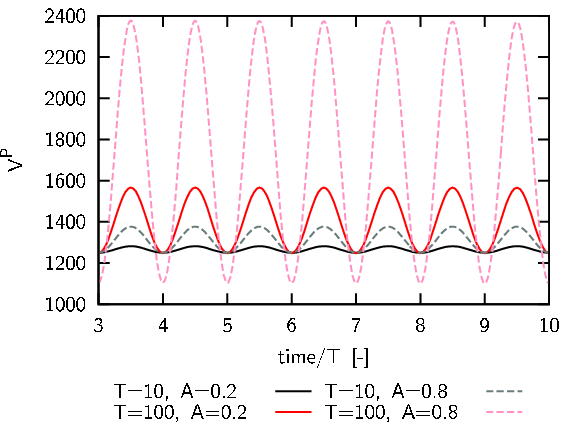}
		\figwithletter{c)}{0.6}{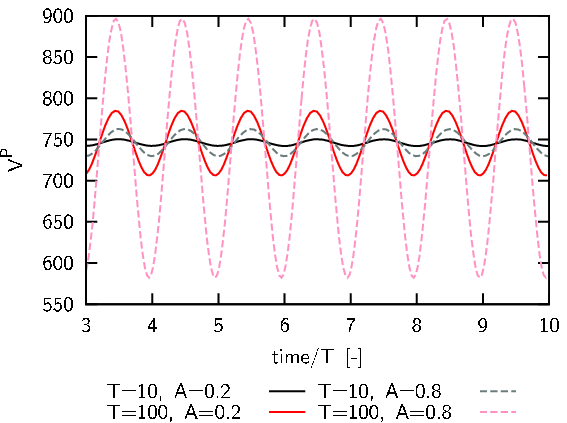}
	\end{center}
	\capt{Fig3SI}{Time traces for regime I (panel \textbf{a}),
          with first-order degradation of $V^P$ and zero-order
          production of $V$ (Eq. 3 of the main text); regime II (panel
          \textbf{b}), with zero-order production and degradation of
          $V$ (\eref{VP2sat}); regime III (panel \textbf{c}), with
          first-order production and degradation of $V$
          (\eref{VP2lin}). In all cases, the amplitude $A$ and
          frequency $T^{-1}$ are varied, while keeping $\mu=50$. The non-linear
          response for the zero-order dynamics in regime II is clearly visible
in
          \textbf{b}. The difference between panel \textbf{a} and
          \textbf{c} is the amplitude of the response. The system of
          the main text with linear production of $V^P$ (I,
          panel \textbf{a}) has a much larger amplitude than that
          with saturated production of $V^P$ (III, panel \textbf{c});
          note the different scale of the y-axis.}
\end{figure*}

\subsection{\slabel{ch8:num_opt}Numerical optimization: a two-step rocket}

\subsubsection{General characteristics}
The numerical optimization used to produce Fig.~5 of the main text is based on
the Wright-Fisher model for population evolution. In each simulation a
population of $N$ independent systems is initialized. Each system consists of a
single multiplexing network.
In the initialization step, each network is assigned random parameters where
each parameter is within the ranges specified in \eref{si:ch8:constraint} and
\eref{si:ranges}.
In the next step the fitness of each network is calculated, which we detail in
the following subsection. Based on the fitness $I_{T}$ for each network, in the
selection step again $N$ new systems are chosen from the original $N$ systems.
The likelihood of selection (reproduction) of each system is proportional to its
fitness $I_T$. Each new network is then ``mutated'' by multiplying all
kinetic parameters by the factor (1 + $\delta$), where $\delta$ is drawn uniform
randomly from the range [−$\Delta$: $\Delta$]; we take $\Delta = 0.3$. Then the
cycle is repeated.

\paragraph{Parameters}\.\newline
We take the kinetic parameters of the encoding module to be fixed, to ensure
correct propagation of the oscillations to component $V$.
For a reliable transmission of oscillations, the
phosphorylation of $V$ is given by Eq. 3 of the main text, as
discussed in the previous section. We further take the mean of the
oscillatory signal to be constant, resulting in the following fixed parameters
\begin{align}
& k_V=1/10\unit{s^{-1}},\nonumber\\
& m_VE_T=5625\unit{s^{-1}},\nonumber\\
& M_V/V_T=30,\nonumber\\
& K_V/V_T=1/250,\nonumber\\
& V_T=2500,\nonumber\\
&\mu_1=25\nonumber\\
\end{align}
Next, the following parameters are constrained based upon values of other
parameters
\begin{align}
\elabel{si:ch8:constraint}
 &k_W \text{ to set } \avg{W^P}=W_T/2 \text{ for a constant signal},\nonumber\\
 &m_{X_2}=m_{X_1}=\br{10T}^{-1},\nonumber\\
 &k_{X_2}=5m_{X_2},\nonumber\\
 &k_{X_1}=X_T m_{X_1}
\br{\frac{W_T^{n_{X_1}}}{W_T^{n_{X_1}}+K_{X_1}^{n_{X_1}}}}^{-1}
\end{align}
The parameters $X_T=W_T=1000$ are constant (unless explicitly mentioned).

This leads to the following set of parameters that are optimized for given
$N_A,N_\mu$, where between square brackets we give the minimum and maximum
value of each parameter.
\begin{align}
\elabel{si:ranges}
&K_W\,\sqbr{1:75000},\nonumber\\
&M_W\,\sqbr{1:75000},\nonumber\\
&m_W\,\sqbr{10^{-3}\,{\rm s^{-1}}:10^3\,{\rm s^{-1}}},\nonumber\\
&T\,\sqbr{10^1\,{\rm s}:10^4\,{\rm s}},\nonumber\\
&\mu_{2,\rm max}\,\sqbr{10:1000},\nonumber\\
&n_{X_1}\,\sqbr{1:5},\nonumber\\
&K_{X_1}\,\sqbr{1:75000}.
\end{align}

\subsubsection{Step 1: Contiguity}
\label{sec:contiguity}
A key point is that, while the precise mapping from $S$ to $X$ may not
be critical for the total amount of information transmitted {\em per
  se}, this is likely to be important for whether or not this
information can be exploited \cite{DeRonde2011}. We therefore impose
the multiplexing requirement \cite{DeRonde2011} (see \fref{si:Fig4}).

\begin{figure*}[!ht]
	\begin{center}
		\figwithletter{}{2.0}{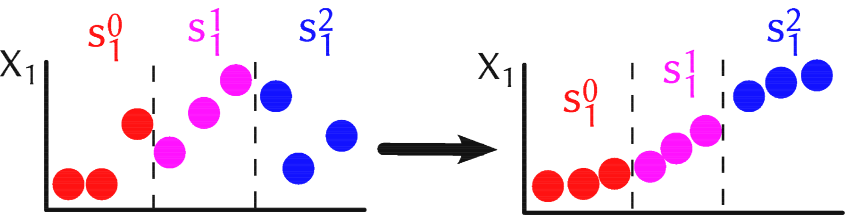}
	\end{center}
   \capt{si:Fig4}{Schematic view of the contiguity requirement. For a system
with $3$ states of $\molsf{S}_1, \molsf{S}_2$, the corresponding output
states of $\molsf{X}_1$ should follow a contiguous (and thus monotonic) order.
In other words, the three states of $X_i$ that correspond to one given state of
$S_i$ are grouped into a set; the different sets $\{X_i\}$ that correspond to
the different states of the input $S_i$ increase monotonically with
$S_i$.}
\end{figure*}

To illustrate the multiplexing requirement,
imagine that each signal $S_i$ can take 3 levels: $S_i^0,S_i^1,S_i^2$
(\fref{si:Fig4}). This means that both
$X_1$ and $X_2$ each have $9$ states, corresponding to the $3\times 3$
possible combinations of input states; for each state of the input
signal $S_i$, {\it i.e.} $S_i^k$, we thus have $3$ output states of $X_i$,
corresponding to
the three different states of the other input signal. The multiplexing
requirement now imposes that the mapping from $S_1,S_2$ to $X_1,X_2$
is such that the output states $\{X_i\}$ corresponding to input
$S_i=j$ are grouped into sets that are contiguous and increase
monotonically with $j$, for each signal $i$. In other words, the three
states of $X_i$ that correspond to one given state of $S_i$ are
grouped into a set; the different sets $\{X_i\}$ that correspond to
the different states of the input $S_i$ increase monotonically with
$S_i$.  This leads to a monotonic input-output relation between $S_i$
and $X_i$ for each $i$, which can be decoded by the network.

Mutual information does not enforce contiguity. Optimizing mutual
information only means minimizing the overlap of the conditional
probability distributions $p(X_i|S_i^k)$ corresponding to the
different states $k$ of the input $S_i$. Certainly in the absence of
noise, where each combination of inputs $S_1,S_2$ yields one and only
one combination of outputs $X_1,X_2$, a minimal overlap of the
conditional distribution does not impose a contiguous division; in
essence, the outputs $X_1$ and $X_2$ are $\delta$-peaks, which could in
principle be arranged in any order when the networks are
optimized for maximizing the mutual information. This, we argue,
hampers decoding. Therefore, the first step in the optimization
routine is to enforce a contiguous mapping.

For a {\it deterministic network} we obtain a contiguous split by the
following procedure: At each step of the evolutionary algorithm we
determine for all input states
($S_1^k,S_2^{k^\prime})=(A_1^k,\mu_2^{k^\prime}$), the output
concentrations $\avgin{X_1}^{k,k^\prime}, \avgin{X_2}^{k,k^\prime}$. For a
specific state $k$ of
the input signal $S_1^k$, $A_1^k$, we determine the minimum and maximum
value of $\avgin{X_1}^{k,k^\prime}$, which (most likely) correspond to
$\avgin{X_1}$ for $\mu_{2, \rm min}$ and $\mu_{2, \rm max}$
respectively. We thus obtain for each state $k$ of $S_1$ a set or
``block'' of $\avg{X_1}$ values between $\acco{\avgin{X_{1, \rm
      min}},\avgin{X_{1, \rm max}}}$ (see also Fig. 4 of the main
text). Similar blocks are obtained for $\avgin{X_2}$ for each state
$k^\prime$ of $S_2$, where the block boundaries depend on $S_1$, {\it
  i.e.} $A_{1, \rm min}, A_{1, \rm max}$.

The fitness of each network is then determined by the amount of
overlap between the different blocks corresponding to the different
states of signal $S_i$, where an increase in overlap
reduces the fitness.
An overlap means that from an
output level $X_i$ the state of the input $S_i$ cannot uniquely be inferred.
Maximum fitness therefore corresponds to minimal overlap between the
blocks.

For a {\it stochastic network}
the optimization method is slightly different. Instead
of determining the boundaries of the block by the minimum and maximum output
concentration, we now include the noise. Using the linear-noise
approximation, we determine  for each output concentration
$\avgin{X_i}$ the corresponding
variance $\sigma^2_{X_i}$. The block is then formed by
$\acco{\sqbr{\avgin{X_{i}}-\sigma_{X_i}}_{\rm
min},\sqbr{\avgin{X_{i}}+\sigma_{X_i}}}_{\rm max}$, where $\sigma_{X_i}$ is the
noise determined through the linear-noise approximation as described in the
main text. The evolutionary algorithm optimizes each network using a minimum
overlap as selection criterion, similar to the deterministic network.

\subsubsection{Step 2: Mutual information}
The procedure outlined above generates networks with optimal
contiguity \cite{DeRonde2011}.  To quantify information transmission in these
networks we compute  the mutual information. Specifically, the
performance measure is the multiplication of
the relative mutual information of the individual channels
\begin{align}
I_T=\frac{I\br{S_1,X_1}}{H\br{S_1}}\times\frac{I\br{S_2,X_2}}{H\br{S_2}},
\end{align}
where $H\br{S_1}$ is the entropy of the amplitude input distribution
$p\br{A_1}$; $H\br{S_2}$ is the entropy of the concentration input
distribution $p\br{\mu_2}$; and $I(S_i,X_i)$ is the mutual information
between $S_i$ and $X_i$ \cite{Shannon1948}.

To determine the entropy of the input distribution and the mutual
information we need to specify the form of the input distributions. We take the
input distributions to be uniform:
\begin{align}
p\br{A_1=a}&=\frac{1}{N_A}, \text{ with amplitude values }
\nonumber\\a_i&=\frac{i}{N_A}, i\in \sqbr{1:N_A},\\
p\br{\mu_2=\mu}&=\frac{1}{N_\mu} \text{ with concentration values
}\nonumber\\
\mu_j&=\frac{j}{N_\mu}\mu_{\rm max}, j\in \sqbr{1:N_\mu},
\end{align}
where $\mu_{\rm max}$ is an optimization parameter with maximum value $1000$.

To compute the mutual information, we have used a slightly different
approach for the stochastic and the deterministic networks.
\paragraph{Mutual information for a stochastic network}
\quad\newline
To calculate the mutual information $I(S_i;X_i)$ for a
stochastic network, we determine for all input
states ($S_1^k,S_2^{k^\prime})=(A_1^k,\mu_2^{k^\prime}$), the output
concentrations $\avgin{X_1}, \avgin{X_2}$ and corresponding variances
$\sigma^2_{X_1},\sigma^2_{X_2}$ via the linear-noise approximation.
With these quantities the mutual information of the two respective channels are
calculated via
\begin{align}
I(S_i,X_i) = H(X_i) - H(X_i|S_i),
\end{align}
where $H(X_i) = -\sum_l p(X^l_i) \ln p(X^l_i)$ and $H(X_i|S_i) = -\sum_k
 p(S_i^k) \sum_l p(X^l_i|S_i^k) \ln p(X^l_i|S_i^k)$.
Here, $p(X^l_i|S_i^k) = \sum_{k^\prime} p(S_{j\neq i}^{k^\prime})
p(X^l_i|S_i^k,S_{j\neq i}^{k^\prime})$, where $p(X^l_i|S_i^k,S_{j\neq
i}^{k^\prime})$ is a
Gaussian distribution centered around the mean $\avgin{X_i}$ given by $S_i^k$
and $S_j^{k^\prime}$.

\paragraph{Mutual information for a deterministic network}
\quad\newline
For a deterministic network without
noise in the mean field limit,  each input $(S_i^k,S_j^{k^\prime})$ maps onto a
unique output $(X^{k,k^\prime}_i,X^{k,k^\prime}_j)$ which is a
delta peak.
One may therefore think that when the number of input
states for each signal goes to infinity, the mutual information also
goes to infinity; this would imply that an infinite number of states
for each signal $S_i$ could be transmitted with 100\% fidelity. However, this is
not true: the mutual information and hence the number of signals that can be
transmitted with 100\% fidelity, remains bounded because of the
crosstalk (and the finite copy number). As described in the text, crosstalk
means that the
input-output mapping no longer is unique; from the output $X_i$, the
input $S_i$ can no longer be inferred with 100\% fidelity.

To compute the maximum amount of information that can be transmitted
through each channel, we adopt the following procedure. We first take
the number of states $N_i$ that is transmitted through each channel
$i$ to be
finite. We thus discretize each signal $S_i$ with equally spaced values
$S_i^k$, with $k=0,\dots, N_i$. Signal $S_1$ is discretized between
$\sqbr{A_{\rm min}-1}$ and signal $S_2$ between $\sqbr{\mu_{\rm min}-\mu_{\rm
max}}$; only $A_{\rm
min}$ and $\mu_{\rm min}$ depend on the number of states; the maximum values
are constant. For each $S_i^k$, we compute $X_i^{k,k^\prime}$ for each state of
the other input signal $S_{j\neq   i}^{k^\prime}$ by solving the mean-field
network in steady state. We then determine the minimum and maximum
values of $X_i^{k,k^\prime}$ for a given $k$,
$X^k_{i,\rm min}$ and $X^k_{i,\rm max}$. This is equivalent to the
``block''-procedure, described above.
Next, we calculate $H(X_i^k|S_i^k)=-\int_{X_{i,\rm
min}^k}^{X^k_{i,{\rm max}}} dX_i^k p(X_i^k|S_i^k)\ln p(X_i^k|S_i^k)$, with
$p(X_i^k|S_i^k)=1/(X_{i,{\rm max}}^k-X_{i,{\rm min}}^k)$ for each
state $k$ of signal $S_i$, $S_i^k$.
To compute $H(X_i|S_i)$, we now have to average $H(X_i^k|S_i^k)$
 over all $S_i^k$:  $H(X_i|S_i) = -\sum_k p(S_i^k)
H(X_i^k|S_i^k)$. The entropy of the output distribution is $H(X_i) =
-\sum_l p(X^l_i) \ln p(X^l_i)$, where $p(X_i) = \sum_k p(S_i^k) p(X_i^k|S_i^k)$.

\fref{si:Fig5}a shows the mutual information $I(S_1;X_1)$ as
function of the of the number of states $N_A$ in channel 1, for
$N_\mu=16$ states of channel 2. It is seen that initially the
mutual information increases linearly with $N_A$; moreover, the slope
is unity. In this regime, the number of states that can be transmitted
with 100\% fidelity through channel 1 is the total number of states of
that channel. In essence, the different blocks of states $X_1$
corresponding to the different states of $S_1$ do not overlap, which
means that from the output $X_1$, the input $S_1$ can be uniquely
inferred. However, as $N_A$ increases beyond 4 bits, the different
blocks overlap increasingly, and the number of signals
that can be transmitted with 100\% fidelity saturates; in the plateau
regime, increasing the
number of input signals further no longer increases the number of
signals that can be transmitted reliably. The plateau
value slightly depends on the number of signals $N_\mu$ that are
transmitted through channel 2, which is shown in panel b. It is seen that the
plateau value saturates to about 3.74 bits when $N_\mu$
is larger than $3.5$ bits. We thus conclude that the maximum
number of signals that can be transmitted with 100\% fidelity through
channel 1 is about 4 bits.

\fref{si:Fig5}c shows the mutual information $I(S_2,X_2)$ as a
function of $N_\mu$, for $N_A=16$. Clearly, the
mutual information is to an excellent approximation given by the
entropy of the input distribution over the full range of $N_\mu$,
which means that all signals can be transmitted with 100\% fidelity,
even when the number of signals goes beyond 4 bits. This is because
the effect of crosstalk from the other channel is negligible, as
explained in the main text.

\begin{figure*}[!ht]
	\begin{center}
		\figwithletter{a)}{0.6}{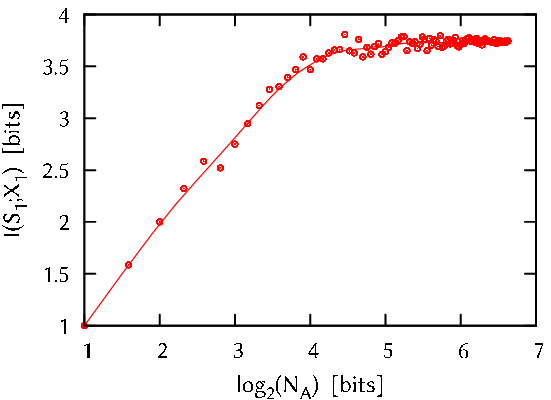}
		\figwithletter{b)}{0.6}{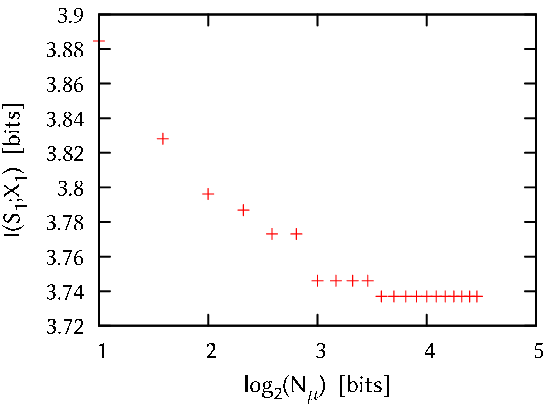}
		\figwithletter{c)}{0.6}{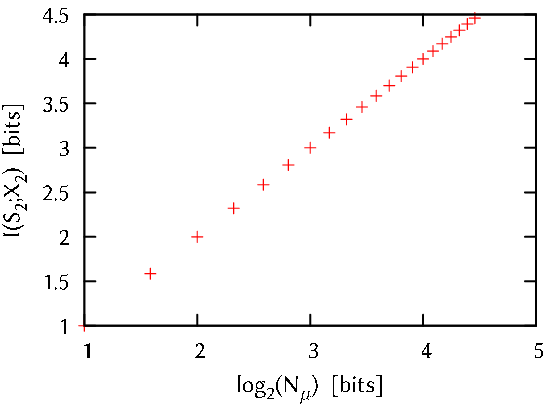}
	\end{center}
        \capt{si:Fig5}{The effect of crosstalk on information
          transmission in a deterministic system. \textbf{a)} The
          mutual information $I(S_1;X_1)$ as function of the number of
          states $N_A$, with $N_\mu=16$. The line is to guide the
          eye. \textbf{b)} The plateau value of $I(S_1;X_1)$ as a
          function of $N_A$ for a given $N_\mu$ (see panel a), plotted
          against $N_\mu$. It is seen that beyond $N_\mu = 3.5$, the
          plateau value is constant.  \textbf{c)} The mutual
          information $I(S_2;X_2)$ as a function of $N_\mu$ for
          $N_A=16$. It is seen that the mutual information is equal to
          the entropy of the input distribution. This is because the
          effect of crosstalk is negligible.  The results are obtained
          for a network that has been optimized via the procedure
          described in section \ref{sec:contiguity} with
          $N_A=N_\mu=16$, and with $A_1$ in the range $\sqbr{0-1}$ and
          $\mu_2$ in the range $\sqbr{0-23}$; the optimized value of
          $\mu_2^{\rm max}=23$ as found by the procedure described
          in \ref{sec:contiguity} is lower than the maximally
          allowed value $\mu_2^{\rm max}=1000$, because that mimizes
          the effect of crosstalk from channel 2 on channel
          1. Parameters:$m_w=0.006{\rm s^{-1}}$, $K_W/W_T=1.6\times
          10^{-2}$, $M_W/W_T=7.9\times 10^{-2}$,
          $K_{X_1}/X_T=1\times10^2$, $n_{X_1}=3$, $X_T=W_T=100$.}
\end{figure*}
\clearpage

\bibliographystyle{plos2009}
\bibliography{FM_multiplex}

\end{document}